\documentclass[a4paper,onecolumn,allowtoday,unpublished]{quantumarticle}
\pdfoutput=1

\usepackage[english]{babel}
\usepackage[letterpaper,top=2cm,bottom=2cm,left=3cm,right=3cm,marginparwidth=1.75cm]{geometry}

\usepackage{amsmath, amsthm, bm, amsfonts, mathrsfs, bbm} 
\usepackage{physics}
\usepackage{graphicx}
\usepackage[x11names,dvipsnames]{xcolor}
\usepackage[colorlinks=true, allcolors=blue]{hyperref}
\usepackage{dsfont}
\usepackage{tikz,pgfplots}\pgfplotsset{compat=1.18}

\def\Pr{\operatorname{Pr}}
\def\Tr{\operatorname{Tr}}

\newcommand{\g}{\mathcal G}
\newcommand{\z}{\mathcal Z}
\newcommand{\zg}{\mathcal Z \circ \mathcal G}

\newcommand{\pkey}{p^{\mathrm{K}}}

\newcommand{\pro}[1]{\ket{#1}\bra{#1}}
\newcommand{\mc}[1]{\mathcal{#1}}

\newtheorem{definition}{Definition}[section]
\newtheorem{proposition}{Proposition}[section]
\newtheorem{condition}{Condition}[section]
\newtheorem{remark}{Remark}[section]

\hypersetup{pdftitle={Finite-size variable-length RPSK}}

\begin{document}

\author[1]{Carlos Pascual-García} \orcid{0000-0003-0659-7349} \thanks{carlos.pascual@luxquanta.com}
\affil[1]{Luxquanta Technologies S.L., Av. Joan Carles I, 30, 1º1ª. 08908 L´Hospitalet de Llobregat, Barcelona, Spain}

\date{}
\title{Finite-size general security for relativistic phase shift keying via variable-length quantum key distribution}

\begin{abstract}
    Differential phase shift keying constitutes a pathway towards practical quantum key distribution by using affordable commercial technologies, and robust theoretical foundations. Recent advances have proven its security against general adversaries, albeit requiring limitations, including strong repetition rate constraints at the security proof and costly statistical estimators. In this work, we overcome said limitations by considering an alternative scheme, herewith denominated relativistic phase shift keying (RPSK). We leverage variable-length general security techniques via entropy accumulation, together with methods based on Rényi leftover hashing and conic optimization. Our approach achieves secret key rates with $10^5$ signals beyond 12 dB, constituting a robust proof of the experimental implementability of RPSK.
\end{abstract}

\maketitle

\section{Introduction}

Quantum key distribution (QKD) \cite{BB84,E91,BBM92, renner2006security} allows any two honest parties, Alice and Bob, to generate a key that is secret with respect to any quantum adversaries (traditionally denominated Eve) \cite{Renner2022Security, ferradini2025definingsecurityquantumkey}. Thanks to its combination of quantum states and classical data postprocessing, it represents the most mature technology for quantum communications, offering information-theoretic security against the incoming threat of cryptographically relevant quantum computers.

In this sense, discrete-variable protocols \cite{pirandola2020advances} allow the implementation of QKD via genuine quantum states, such as single photons and weak coherent pulses. Specifically, decoy-state BB84 \cite{Lo2005,kamin25MEATsecurity} has become the main platform for the design and certification of QKD standards \cite{tupkary2026rigorouscompletesecurityproof} thanks to its high yield and large communication distances. However, discrete-variable QKD comes at the cost of using expensive specific hardware, particularly single-photon detectors, as well as repetition rate limitations due to deadtimes. 

On the other hand, continuous variable QKD (CVQKD) \cite{grosshans2002continuous, zhang2024continuousvariable} offers an affordable approach as it is based on classical protocols that are already under use in telecommunications, together with using commercial technologies -- thereby representing a scalable path towards the practical deployment of QKD. Nevertherless, long-distance CVQKD under general security still poses multiple challenges. From a practical perspective, its low signal to noise ratios and excess noise levels \cite{leverrier2023information} require the development of new error correction methods to decrease the time overhead of the classical postprocessing. From a theoretical point of view, its complex framework based on infinite-dimensional states  often requires either costly, de Finetti-type symmetrizations \cite{Leverrier2015} or dimension reduction techniques \cite{Upadhyaya2021Cutoff,Kanitschar2023,primaatmaja2024,navarroPostSel2026} which require a numerically intensive optimization.

Within this context, differential phase shift keying (DPSK) \cite{Inoue2002,Takesue_2005,Mizutani_2024Asymptotic,Mitzutani2024,Sandfuchs2025} has gained notorious attention due to its optimal tradeoff by using conventional, optical devices while offering critical transmittances at the same level as discrete-variable QKD. In contrast with its discrete variable counterparts, it avoids the well-known photon number splitting attack \cite{Huttner1995,Brassard_2000} thanks to its encoding of the information between measurements. Compared to continuous variable QKD, it needs more expensive equipment such as photodetectors, but it exploits the no-clicks by Bob as a postselection tool, thereby allowing DPSK to outperform any critical transmittances that can be achieved by CVQKD.

Recently\footnote{See also Ref. \cite{Mitzutani2024} for a general security proof without repetition rate restrictions, albeit at the cost of a suboptimal key rate per pulse, and a detection based on photon number resolution.}, Ref. \cite{Sandfuchs2025} has shown that DPSK allows the distribution of secret keys beyond 30 dB (around 150 km of commercial optical fibers) while enforcing general security. However, the scope of the security proof relies on sequential limitations from the generalized entropy accumulation theorem (GEAT) \cite{metger2022security}, which fundamentally limits the repetition rate of the protocol to the order of kHz while also requiring sharply large block sizes. This constraint persists when using more advanced entropy accumulation techniques, as these require a single-round analysis and do not include the memory registers needed for DPSK, where signals and measurements are interwinded in consecutive rounds.

With this work, we overcome said limitations by focusing on the relativistic variant presented in \cite{Sandfuchs2025}, herewith denoted as relativistic phase shift keying (RPSK) for short, whose structure allows a single-round description and is therefore amenable for a finite-size analysis based on the marginal-constrained entropy accumulation theorem (MEAT) \cite{arqand2025MEAT}. In particular, we use a variable-length framework \cite{Tupkary2024,kamin25MEATsecurity} for the key analysis which guarantees a resilience for the protocol implementation against imprecise characterizations of the quantum channel. Our results show an improvement for the block size in orders of magnitude, together with  an analysis of experimental imperfections, including the cost of dark counts in the secret key, as well as an asymmetric modulation for Alice's quantum states.

The rest of this document goes as follows. Section \ref{Sec:Preliminaries} provides diverse mathematical concepts of interest. Section \ref{Sec:ProtocolDescription} follows the description of the RPSK protocol from \cite{Sandfuchs2025}, together with some technical derivations that are relevant for the security analysis. Then, Section \ref{Sec:VarLen} provides the technical framework of the general security proof via MEAT, together with the numerical optimization tools required to derive the secret key rate. Given this characterization, several results are presented in Section \ref{Subsec:NumResults} to illustrate the performance of RPSK, whose impact we discuss in Section \ref{Sec:Discussion}.

\section{Mathematical prolegomena} \label{Sec:Preliminaries}

Let $\mc{H}_X$ denote a Hilbert space for a register $X$, and denote $\mc{D}(X)$ as the space of all density matrices defined on said register. We define $X_1^n = X_1 ... X_n$ for $n$ identical copies whose embedding is provided by $X$. Let any bipartite state $\rho_{CX}$ be a classical-quantum state (cq-state) whenever register $C$ is classical and the state is therefore represented as
\begin{equation}
    \rho_{CX} = \sum_{c\in \mathcal{C}} p(c) \ketbra{c}{c}_C\otimes \rho^c_X,
\end{equation}
for some alphabet $\mc{C}$. For a subset $\Omega \subseteq \mc{C}$, we denote the conditioned state as
\begin{equation}
    \rho_{CX|\Omega} = \frac{1}{\Pr[\Omega]_{\rho}}\sum_{c\in \Omega} p(c) \ketbra{c}{c}_C\otimes \rho^c_X,
\end{equation}
where $\Pr[\Omega]_{\rho} = \sum_{c \in \Omega} p(c)$. We further define the $f-$weighted Rényi entropy \cite{arqand2024generalized,arqand2025MEAT} for a cqq-state $\rho \in \mc{D}(CXY)$ classical in $C$ and $\alpha \in (0,\infty)\backslash \{1\}$ as 

\begin{align}
     H^{\uparrow,f}_\alpha (X|{C}Y)_{\rho} &= \frac{\alpha}{1-\alpha} \log \left( \sum_{{c}\in \mathcal{C}} p({c}) 2^{\frac{1-\alpha}{\alpha} (-f ({c})+ H^\uparrow_\alpha (X|Y))_{\rho_{|{c}}}}  \right) 
\end{align}
where $f: \mc{C} \to \mathbb{R}^{|\mc{C}|}$ is called a tradeoff function (whose choice can be arbitrary), and 
\begin{equation}
    H^\uparrow_{\alpha}(X|Y)_\rho = \sup_{\sigma_Y \in \mc{D}(Y)} - D_\alpha(\rho_{XY}\|\mathds{1}_X \otimes\sigma_Y),
\end{equation}
is the sandwiched Rényi entropy \cite{tomamichel2015quantum}, expressed via the sandwiched Rényi relative entropy
\begin{equation} \label{eq:QRenyiDiv}
    D_\alpha(\rho\|\sigma) = \frac1{\alpha-1}\log\left(\frac{\Psi_\alpha(\rho, \sigma)}{\Tr[\rho]}\right),
\end{equation}
where 
\begin{equation}\label{eq:PsiRenyi}
    \Psi_\alpha(\rho, \sigma) = \Tr\left[(\sigma^\frac{1-\alpha}{2\alpha}\rho\sigma^\frac{1-\alpha}{2\alpha})^\alpha\right].
\end{equation}
To conclude, we define the relative entropy 
\begin{equation}
    D(\rho||\sigma) = \Tr[\rho \log(\rho) - \rho \log(\sigma)],
\end{equation}
which becomes the Kullback-Leibler divergence $D_{\mathrm{KL}}(\cdot || \cdot)$ for two classical distributions.

\section{Protocol outline}\label{Sec:ProtocolDescription}

As a first step in our analysis, we start by formulating the relativistic QKD protocol from \cite{Sandfuchs2025} under the framework provided by the MEAT \cite[Section 6]{arqand2025MEAT}. In particular, we highlight the relativistic constraints that are needed in order to enforce security under our framework, eventually allowing a characterization of the secret key rate that can be achieved. 

RPSK is based on Alice sending two states in every round at different times. The first one is a reference, given by a coherent state whose phase and amplitude are fixed, and the second one is a coherent state of the same amplitude but unknown phase. Bob extracts the information encoded by Alice via correlations between the two quantum states and threshold detection, while also verifying that their corresponding times of arrival coincide with their expected timestamps. As a result, the time tags set by Alice and Bob play a crucial role in the characterization of the protocol -- we define, for every round $i \in [n]$, the time-ordering

\begin{subequations}\label{eq:TimeOrderingRPSK}
\begin{align}
    {t}^{A}_{i} &< {t}^{A'}_{i} \\
    {t}^{B}_{i} &< {t}^{B'}_{i} \\
    {t}^{B'}_{n} &< {t}^{\mathrm{ann}}_{1} \leq \dots \leq {t}^{\mathrm{ann}}_{n}
\end{align}
\end{subequations}

Where
\begin{itemize}
    \item ${t}^{A}_{i}, {t}^{A'}_{i}$ are time tags for Alice sending the reference and signal states to Bob, forming each a monotonically increasing sequence.
    \item $ {t}^{B}_{i}, {t}^{B'}_{i}$ are the expected times of arrival at Bob's laboratory for the reference and signal states, satisfying the relation $ {t}^{B}_{i} =  {t}^{A}_{i} + v/d$ (idem for $A'$ and $B'$) where $d$ is the distance separating Alice and Bob, and $v$ is the speed of light at the channel. 
    \item ${t}^{\mathrm{ann}}_{i}$ is the time of public announcements by Alice and Bob corresponding to round $i$.
\end{itemize}

First of all, let us note that this structure does not use on-the-fly announcements \cite{tupkary2026rigorouscompletesecurityproof}, which allows a simplified security analysis since the maps representing every round can be assumed to be in a tensor product \cite{arqand2025MEAT}. 

The time-tagging for the signals will allow us to constrain Eve's attack to local interactions between the signal and reference states, as provided in Subsection \ref{subsec:Nosignaling}. In particular, Bob receives the two quantum states for a given round at different times (which amounts to a time-bin encoding) and his joint measurement requires to correlate the reference and the signal. This is done, for instance, by following the approach of Ref.  \cite[Fig. 21]{Sandfuchs2025} where Bob applies a delay in the reference, already present in his laboratory, to make it coincide with the signal at a beam splitter.

\begin{remark}
    Note that we do not impose any conditions over the times of arrival between the two different signals of Alice and Bob, even for the same round, particularly of the kind $t_i^B < t^{A'}_i$. Albeit the ascending ordering of the timestamps for the rounds eventually limit Eve's ability to generate correlations between signals, it nevertheless allows her to perform an indirect attack (e.g. using entangled pairs) over any two or more signals that are present in the quantum channel. Hence, the physical constraints imposed by the no-signaling as provided later in Sec. \ref{subsec:Nosignaling} do not rule out the possibility of a general attack.
\end{remark}

Provided these considerations for the ordering of their actions, Alice and Bob perform the protocol as follows.
\begin{enumerate}
  \item \textbf{Preparation and measurement.} For $i\in[n]$:
  \begin{enumerate}
    \item Alice draws a bit $V_{i}\in\{0,1\}$ randomly, prepares a reference $\ket{\beta}_{\tilde{R}}$ and a signal $\ket{(-1)^{V_{i}}\beta}_{\tilde{S}}$ and sends them to Bob at different times such that no-signaling\footnote{This means that the quantum signals do not coincide in any region of the spacetime, such that Eve cannot make them interact directly without causing an abortion for the protocol.} between the states is enforced.
    \item Bob receives the states $\rho_{S}$ and $\rho_R$, makes them interact via a beam splitter, and performs a threshold detection given by the POVM $\{M^{(b)}_{S'R'}\}_{b}$. He records his measurement outcome in a register $B_{i}\in\{0,1,\perp\}$. In the case of a double-click, he randomly reassigns the round as $B_i \in \{0,1\}$. 
    \item If the outcome is a click (i.e., $B_i = 0,1$), Bob draws a bit $I_i$ with probabilities $(\pkey,1-\pkey)$. In the case of a no-click, he sets $I_i=1$.
  \end{enumerate}
  \item \textbf{Public announcement.} Bob announces the value of $I_1^n$, and then Alice and Bob reveal their values $V_i B_i$ from all rounds such that $I_i=1$. With said announcements, Alice and Bob build a common register $C_1^n$. Alice further sets the raw key register $Z_i$ with $Z_i = V_i$ whenever $I_i = 0$, and $Z_i=\perp$ otherwise.
  \item \textbf{Postprocessing.} Using $C_1^n$ Alice and Bob bound Eve's information and perform a variable-length decision, which determines the size of the final secret key. In particular, if the discussion reveals any mismatch between the expected and the actual time tags, the key length is set to zero and the protocol concludes.
  \item \textbf{Information reconciliation} Based on $C_1^n$, Alice sends $\lambda_\mathrm{EC}(c_1^n)$ bits of classical information about her key, which Bob uses together with $B_1^n$ to derive a guess of Alice's secret key. The process is validated by Alice and Bob sharing a universal$_2$ hash (at the cost of revealing $\lceil \log(1/\varepsilon_\mathrm{EC}) \rceil$ secret bits). If their hashes do not coincide, the length of the final secret key is set to zero and the protocol concludes.
   \item \textbf{Privacy amplification} Alice and Bob remove any remaining side information by Eve about their shared keys by using another universal$_2$ hash, which results in the final secret key.
\end{enumerate}

Provided this description, let us explicitly define register $C$, whose alphabet will be composed by key generation rounds ($\perp$), correct clicks ($\mathrm{CC}$), wrong clicks ($\mathrm{WC}$) and no-clics ($\mathrm{NC}$). Hence, we define the set
\begin{equation}
    \mc{C} = \{\perp, \mathrm{CC}, \mathrm{WC}, \mathrm{NC}\}.
\end{equation}
On similar grounds, we will use the restriction $\tilde{\mc{C}}:= \mc{C} \backslash \{\perp\}$ for the non-trivial symbols announced during the classical postprocessing. These values are built from Bob's measurement outcomes, where the POVM is modeled via threshold detection \cite{Sandfuchs2025}

\begin{subequations}\label{eq:ClickPOVM}
\begin{align}
    M^\bot_{S'R'} &= \pro{0,0} \\
    M^0_{S'R'} &= 
\sum_{N=1}^{\infty} \lvert N,0\rangle \langle N,0\rvert  \\
    M^1_{S'R'} &= 
\sum_{N=1}^{\infty} \lvert 0,N\rangle \langle 0,N\rvert  \\
M^\mathrm{dc}_{S'R'} &=
\sum_{N=2}^{\infty} \sum_{n=1}^{N-1} 
\lvert N-n, n\rangle \langle N-n, n\rvert. 
\end{align}
\end{subequations}
These POVM elements describe respectively the events of no-click, and clicks on the detectors zero, one and both (double-click). The rationale for this detection lies on the fact that the coherent states received by Bob create a well-defined pattern of clicks after the beam splitting. In the case of an ideal, lossless channel, Bob applies the beam splitter action $U_\mathrm{bs}$ on the received states and observes the two possible outcomes
\begin{align}
U_\mathrm{bs} \ket{\beta}_{{S}} \otimes \ket{\beta}_{{R}} &= \ket{\sqrt{2}\beta}_{S'} \otimes \ket{0}_{R'} \\ 
U_\mathrm{bs} \ket{-\beta}_{{S}} \otimes \ket{\beta}_{{R}} &=  \ket{0}_{S'} \otimes \ket{-\sqrt{2}\beta}_{R'}. 
\end{align}
Such that Bob can maximally recover the information encoded by Alice via the clicks at his two detectors, as modeled with the POVM \eqref{eq:ClickPOVM}.

Next, let us consider a more realistic model for Bob's detection by including dark counts on the measurements. Furthermore, we consider that Bob randomly assigns double clicks to zero or one. Provided the dark count probabilities $p_d^0, p_d^1 \in [0,1]$ for each detector, the POVM that incorporates these considerations is achieved by a linear transformation \cite{Nahar_2026}
\begin{align} \label{eq:DarkCounts}
    \begin{pmatrix}
        \bar{M}^\bot \\ \bar{M}^0 \\ \bar{M}^1 
    \end{pmatrix} = 
    \begin{pmatrix}    
    (1-p_d^0)(1-p_d^1) & 0 & 0 & 0  \\
    p_d^0(1-p_d^1/2) & 1-p_d^1/2 & p_d^0/2 & 1/2 \\
    p_d^1(1-p_d^0/2) & p_d^1/2 & 1-p_d^0/2 & 1/2
    \end{pmatrix} \cdot \begin{pmatrix}
        M^\bot \\ M^0 \\ M^1 \\ M^\mathrm{dc}
    \end{pmatrix}
\end{align}
For the sake of simplicity, we will assume a symmetric implementation where $p_d^0 = p_d^1 = p_d$ from here onwards. This is an assumption that can be done without loss of generality, as dark counts constitute a trusted noise and Bob can always add more artificial noise at the detectors to make them match.  

To conclude, we note that the measurement is actually applied on the states after these pass through Bob's beam splitter. Accordingly, one can rather absorb the action of said beam splitter into Bob's detection, and directly apply the transformed POVM on the incoming states. For later convenience, let us consider the adjoint action $U^\dagger_\mathrm{bs}$ of the beam splitter over the POVM. We start from the relation \cite[Eq. (70)]{Sandfuchs2025} 
\begin{align}
    U^\dagger_\mathrm{bs} \left( {M}^{0} - M^1 \right)U_\mathrm{bs} &= 
    \sum_{N=1}^\infty \frac{1}{2^N} \sum_{m,n=0}^{N} \sqrt{\begin{pmatrix}
    N \\ m
\end{pmatrix}} \sqrt{\begin{pmatrix}
    N \\ n
\end{pmatrix}}[1 - (-1)^{m+n}]\ket{N-m,m}\bra{N-n,n}.
\end{align}
Provided the click operators of \eqref{eq:DarkCounts}, we leverage the equation above to find

\begin{align}\label{eq:AfterBS}
    U^\dagger_\mathrm{bs} \bar{M}^{0/1} U_\mathrm{bs} &= \frac{1}{2} \mathds{1} - \frac{(1-p_d)^2}{2} \pro{00} \pm \frac{1-p}{2} U^\dagger_\mathrm{bs} (M^0 - M^1) U_\mathrm{bs} \\
    & \quad\frac{1}{2} \mathds{1} - \frac{(1-p)^2}{2}\pro{00} \nonumber \\
    & \quad \pm \frac{1-p}{2} \sum_{N=1}^\infty \frac{1}{2^N} \sum_{m,n=0}^{N} \sqrt{\begin{pmatrix}
    N \\ m
\end{pmatrix}} \sqrt{\begin{pmatrix}
    N \\ n
\end{pmatrix}}[1 - (-1)^{m+n}]\ket{N-m,m}\bra{N-n,n}, 
\end{align}
where we exploit the fact that the no-click and identity operators are invariant under the beam splitting. 

\subsection{Source and squashing maps}\label{subsec:Maps}

Crucially, the MEAT is only well-defined for finite-dimensional states \cite{arqand2025MEAT,tupkary2026rigorouscompletesecurityproof}. Therefore, it cannot be readily implemented in protocols that require optical states. This problem can be overcome by using source maps at Alice's total preparation, as well as squashing maps at Bob's measurements. Both tools reduce the dimensionality of all registers required by the MEAT into a compact support, allowing among others the implementation of QKD using coherent states.

In the case of Alice's preparation, we note that it is provided by pure, rank-one states $\{\ket{\alpha}, \ket{-\alpha}\}$. Hence, Eve receives quantum signals whose span is finite so that we can restrict without loss of generality her input to this two-state subspace. The next task is to perform a dimension reduction for Bob's received states. 

This can be done by using a squashing map \cite{Beaudry_2008,Gittsovich_2014}, which acts on the individual registers of Bob by reducing the measurement on a large, possibly infinite-dimensional space into a finite, compact support while preserving the observed outcomes. Given these conditions on the registers held by Alice and Bob, it is then possible to show that Eve's optimal attack is to use finite-dimensional registers \cite[Appendix A]{tupkary2026rigorouscompletesecurityproof}.

\begin{definition}[\textbf{Squashing map}] \emph{\cite[Definition 2.13]{Sandfuchs2025} }\label{Def:Squashing}
Let $A$ and $A'$ two quantum registers, and $\{M^x_A\}_{x \in \mc{X}}$, $\{N^x_{A'}\}_{x \in \mc{X}}$ two POVMs for said registers. A CPTP map $\Lambda:A' \leftarrow A$ is a squashing map from  $\{M^x_A\}_{x \in \mc{X}}$ to $\{N^x_{A'}\}_{x \in \mc{X}}$ if for all $x \in \mc{X}$ and all $\rho \in \mc{D}(A)$ it verifies
\begin{equation}\label{eq:SquashingMap}
    \Tr[M^x_A \rho] = \Tr[N^x_{A'} \Lambda(\rho)].
\end{equation}  
\end{definition}
We note that, in principle, the squashing is a virtual operation applied by Bob at his laboratory. Hence, we can fully characterize it via a map $\Lambda(\cdot)$, give $\Lambda(\cdot)$ to Eve (since this can only increase the strength of her attack) and consider that Bob directly applies his measurements via the POVM  $\{N^x_{A'}\}_{x \in \mc{X}}$ on the state after the squashing. As a result, we may absorb the corresponding squashing map and directly consider $\omega_{SR} = \Lambda(\rho_{SR})$ for the following analysis where, with a slight abuse of notation, we assume that this new state lives in the same registers $SR$.

Within our particular context, we may employ the squashing from Ref. \cite[Theorem 2.14]{Sandfuchs2025}, which also provides the squashed POVM for the threshold detection in the case of $p_d = 0$

\begin{subequations}\label{eq:squashPOVM}
\begin{align}   
N^{0}_{SR} &= \lvert \phi^{+} \rangle \langle \phi^{+} \rvert 
+ \frac{1}{2} \lvert 11 \rangle \langle 11 \rvert \\ 
N^{1}_{SR} &= \lvert \phi^{-} \rangle \langle \phi^{-} \rvert 
+ \frac{1}{2} \lvert 11 \rangle \langle 11 \rvert \\ 
N^{\perp}_{SR} &= \lvert 00 \rangle \langle 00 \rvert,
\end{align}
\end{subequations}
where $\ket{\phi^\pm} = [\ket{01} \pm \ket{10}]/\sqrt{2}$. Furthermore, we will define the click operator as $N^\top := N^0 + N^1$, which shall become useful in order to define the key distillation map. 

In the case of adding dark counts to the detection, \eqref{eq:squashPOVM} can be generalized following the same principle as in the original POVM. For the no-click operator, the addition is a simple re-scaling, whereas a noisy click operator from \eqref{eq:DarkCounts}, together with the action of the beam splitter as in \eqref{eq:AfterBS} provides

\begin{align}
    \Tr[U^\dagger_\mathrm{bs} \bar{M}^{0/1} U_\mathrm{bs}\rho] &= \frac{1}{2} \Tr[\rho] - \frac{(1-p_d)^2}{2} \Tr[U^\dagger_\mathrm{bs} M^\bot U_\mathrm{bs} \rho] \pm \frac{1-p}{2} \Tr[U^\dagger_\mathrm{bs}(M^0 - M^1) U_\mathrm{bs} \rho] \\
    &= \frac{1}{2} \Tr[\Lambda(\rho)] - \frac{(1-p_d)^2}{2} \Tr[N^\bot \Lambda(\rho)] \pm \frac{1-p}{2} \Tr[(N^0 - N^1) \Lambda(\rho)] \\
    &=: \Tr[\bar{N}^{0/1} \omega].
\end{align}
The first line uses the decomposition in \eqref{eq:DarkCounts}, the second employs \eqref{eq:SquashingMap} together\footnote{Observe also that $M^0 - M^1 = (M^0 + M^\mathrm{dc}/2) - (M^1 + M^\mathrm{dc}/2)$.} with the trace-preserving nature of the squashing map $\Lambda(\cdot)$, and the third one defines the new POVM element. The definition for the click operator $\bar{N}^\top_{RS}$ follows immediately. 

\subsection{No-signaling}\label{subsec:Nosignaling}

One further condition given by RPSK with respect to standard QKD protocols is provided by a notion of no-signaling between the two states sent by Alice in every round. Namely, if Alice and Bob synchronize their actions and thanks to the limiting factor of the speed of light, Eve is unable to make the two states interact within the same spacetime event without causing an abortion, as the interaction would induce a distortion with respect to Bob's expected times of arrival. In other words, this can be expressed via the following condition.

\begin{condition}\emph{\cite[Condition 3.1]{Sandfuchs2025}}\label{cond:nosignaling}
    Eve does not signal from the signal state to the reference state.
\end{condition}
Using \cite[Eq. (16)]{Sandfuchs2025} and the subsequent derivation, the no-signaling can be recast as a constraint on Eve's attack channel $\mc{E}: SR \leftarrow \tilde{S}\tilde{R}$ as
\begin{equation} \label{eq:Nosignaling}
    \mathrm{tr}_{S} \circ \mathcal{E}_{S R \leftarrow \tilde{S}\tilde{R}} = \mathrm{tr}_{\tilde{S}} \otimes \mathcal{E}_{R \leftarrow \tilde{R}}.
\end{equation}

\begin{remark}\label{rem:enforceNS}
    This condition can be enforced for any quantum channels whose speed is underluminal, provided a set of extra conditions on the time tagging of Alice and Bob -- which ensures that Eve cannot accelerate the reference and/or decelerate the signal to overlap them in a spacetime region (which breaks Condition \ref{cond:nosignaling}). 
    
    As pointed in \cite[Section 5.1]{Sandfuchs2025}, for every $i-$th round we must require that the time of arrival for the reference, as if transmitted in vacuum $\bar{t}_i^B = t_i^{A} + d/c$, where $d$ is the distance between Alice and Bob, must remain larger than the expected time of arrival for the signal $t_i^{B'} = t_i^{A'} + d/v$ where $v$ is the speed of light in the channel. As a result, this condition $ \bar{t}_i^B  \geq t_i^{B'}$ induces the constraint in the implementation
    \begin{equation}\label{eq:OpticalCondition}
        t_i^{A} - t_i^{A'} \geq d \left( \frac{1}{v} - \frac{1}{c}  \right).
    \end{equation}
    Albeit this constitutes a constraint on the repetition rate, we note that it is purely dependent on the quantum channel, and not the security proof. Therefore, it might have impact in the case of optical fibers ($v \approx 2 c/3$) but its effect for free-space and satellite links ($v \approx c$) is nearly negligible.
\end{remark}

In our particular implementation, we have that Alice prepares, in an entanglement-based representation\footnote{This image is fully equivalent to its prepare-and-measure version thanks to the source-replacement scheme \cite{BBM92}.} \cite{E91,BBM92},
\begin{equation} \label{eq:EBpicture}
    \ket{\psi} = \frac{1}{\sqrt{2}}\sum_{j=0,1} \ket{j}_A \otimes \ket{(-1)^j \beta}_{\tilde{S}} \otimes \ket{\beta}_{\tilde{R}}.
\end{equation}
This state evolves according to Eve's attack channel
\begin{align}
    \mathrm{id}_A \otimes \mathcal{E}(\psi) &= \frac{1}{2} \sum_{j,k=0,1} \ket{j}\bra{k}_A \otimes \mathcal{E}\left(  \ket{(-1)^j \beta} \bra{(-1)^k \beta}_{\tilde{S}} \otimes \ket{\beta} \bra{\beta}_{\tilde{R}}\right) \\
    &:= \omega_{A S R}.  \label{eq:FinalState}
\end{align}
Adding the no-signaling condition from \eqref{eq:Nosignaling}, said representation translates to the constraint

\begin{align}
    \Tr_{S}[\omega_{A S R}] &= \frac{1}{2} \sum_{j,k=0,1}   \bra{(-1)^k \beta} \ket{(-1)^j \beta}  \ket{j}\bra{k}_A \otimes \mathcal{E}\left( \ket{\beta} \bra{\beta}_{\tilde{R}}\right) \\
    &:= \sigma_A \otimes \omega_{R},
\end{align}
where we directly denoted $\sigma_A$ as Alice's marginal
\begin{align}
\sigma_A = \frac{1}{2} \sum_{j,k=0,1}  \bra{(-1)^k \beta} \ket{(-1)^j \beta}  \ket{j}\bra{k}_A,
\end{align}
which constitutes an intrinsic constraint in the implementation of the MEAT.


\subsection{Single-round channel} \label{subsec:SingleRoundChannel}

Let us now formulate the map that describes one round of the protocol, according to its description and applying the squashing map at Bob's registers. In particular, we note that the values of $Z$ and $C$ together deterministically provide the value for $I$, such that we remove $I$ without loss of generality from the following expressions. Using an entanglement-based image, where Alice holds a quantum register $A$, and her preparation is provided by \eqref{eq:EBpicture},
\begin{align}
    \mathcal{M}(\omega_{ASR}) =& \pkey \sum_{x=0,1} \Tr[\ketbra{x}_A \otimes \bar{N}^\top_{SR}  (\omega_{ASR})]\ketbra{x}_{Z} \otimes \ketbra{\perp}_C \nonumber \\
    & + (1 -\pkey) \sum_{x=0,1} \Tr[\ketbra{x}_A\otimes \bar{N}^x_{SR} (\omega_{ASR})] \ketbra{\bot}_{Z} \otimes \ketbra{\mathrm{CC}}_C  \nonumber\\
    & + (1 -\pkey) \sum_{x=0,1} \Tr[\ketbra{x}_A\otimes \bar{N}^{\tilde{x}}_{SR} (\omega_{ASR})] \ketbra{\bot}_{Z} \otimes \ketbra{\mathrm{WC}}_C \nonumber \\
    & + \Tr[\mathds{1}_A \otimes \bar{N}^\bot_{SR}  (\omega_{ASR})]\ketbra{\perp}_{Z} \otimes  \ketbra{\mathrm{NC}}_C, \label{eq:SingleRoundMap} \\
    =& \pkey \mc{M}^\mathrm{K}(\omega_{ASR})_{Z} \otimes \pro{\bot}_C + \pro{\bot}_{Z} \otimes \mc{M}^\mathrm{T}(\omega_{ASR})_{C}, \label{eq:SingleRoundRPSK}
\end{align}
where $\tilde{x} = 1 \oplus x$, and $\omega_{ASR}$ is the state shared by Alice and Bob after passing through the (no-signaling) quantum channel, as given by \eqref{eq:FinalState}.  In particular, we note that all rounds resulting in a no-click will be handled as public information, such that the key map $\mc{M}^\mathrm{K}(\cdot)$ is trace-nonincreasing. For the sake of simplifying the numerical description, we change the notation for the combined POVM elements used by Alice and Bob in parameter estimation to
\begin{subequations}\label{eq:CompactOperators}
\begin{align}
    \Gamma^\mathrm{CC} &= (1-\pkey) \sum_{x=0,1} \ketbra{x}_A\otimes \bar{N}^{{x}}_{SR} \\
    \Gamma^\mathrm{WC} &= (1-\pkey) \sum_{x=0,1} \ketbra{x}_A\otimes \bar{N}^{\tilde{x}}_{SR}  \\
    \Gamma^\mathrm{NC} &= \mathds{1}_A \otimes \bar{N}^\bot_{SR},
\end{align}
\end{subequations}
where Bob's POVM may include noise due to dark counts, as explained in Subsec. \ref{subsec:Maps}.

\section{Variable-length general security}\label{Sec:VarLen}

Let us now characterize the length of the secret key that can be achieved under a variable-length, MEAT-based security proof for the RPSK protocol. For this task, we employ \cite[Theorem 12]{kamin25MEATsecurity}, which can be readily adapted to our particular context.

\begin{proposition}\label{Prop:MainProp}
    Let $\alpha \in (1,2)$, $\varepsilon_{\mathrm{PA}}, \varepsilon_{\mathrm{EC}} \in (0,1]$, and 
any tradeoff function $f : {\mathcal{C}} \to \mathbb{R}$. 
Let further $\kappa$ be the ${H}^{\uparrow}_{\alpha, f}$-normalization constant corresponding to the set of all states that can be produced in a single round via a map $\mc{M}: ZCE \leftarrow ASR$, which enforces the no-signaling Condition \ref{cond:nosignaling}. I.e.,

\begin{align}\label{eq:KappaProp}
  \kappa
:= \inf_{\substack{\omega_{ASR}\in\mathcal{D}(ASR)\\
\text{s.t. }\Tr_{S}[\omega_{ASR}]=\sigma_{A} \otimes \omega_{R}}}
\; {H}_{\alpha}^{\uparrow,f} (Z | {C}E )_{\mc{M}(\omega)}.
\end{align}
Then, the RPSK protocol is, provided appropriate source and squashing maps, $\varepsilon_{\mathrm{PA}}$-secret and
$\varepsilon_{\mathrm{EC}}$-correct, producing a secret key whose variable length $\ell$ is, conditioned on a successful error correction, chosen as some function of ${c}_1^n$ satisfying
\begin{align} \label{eq:SKR_Proposition}
    \ell \leq \max \left\{0, \hat{f}({c}_1^n) - \lambda_\mathrm{EC}({c}_1^n) - \left\lceil\log\frac{1}{\varepsilon_{\mathrm{EC}}}\right\rceil-\frac{\alpha}{\alpha-1}\log\frac{1}{\varepsilon_{\mathrm{PA}}}+2\right\}
\end{align}
where $\hat{f}$ is the full tradeoff function
\begin{equation}
    \hat{f}({c}_1^n) = \sum_{c \in \mc{C}} (f(c_i) + \kappa),
\end{equation}
and $\lambda_\mathrm{EC}({c}_1^n)$ is the total amount of raw key bits leaked during the error correction.
\end{proposition}
This Proposition is only valid under protocols that employ finite-dimensional registers as it is based on the MEAT \cite{tupkary2026rigorouscompletesecurityproof}. As provided in Subsection \ref{subsec:Maps}, Alice's preparation is given by a finite span, and we can reduce the dimensionality of Bob's states into a compact support via the squashing map from \cite{Sandfuchs2025}, which defines a virtual protocol with the same security as the corresponding real protocol.

Now, let us expand the minimization for $\kappa$ into quantities that are amenable for numerical analysis. Noting that the protocol is described (after tracing out register $E$) by the single-round map from \ref{subsec:SingleRoundChannel} and using the classical conditioning for sandwiched Rényi entropies \cite{tomamichel2015quantum}, we can express the f-weighted Rényi entropy as \cite{navarro2025}
\begin{align}
    H^{\uparrow,f}_\alpha (Z|{C}E)_{\mc{M}(\omega)} &= \frac{\alpha}{1-\alpha} \log \left( \sum_{{c}\in \mathcal{C}} p(C={c}) 2^{\frac{1-\alpha}{\alpha} \left[-f ({c})+ H^\uparrow_\alpha (Z|E)_{\mc{M}(\omega)_{|{c}}}\right]}  \right)  \\
    &= \frac{\alpha}{1-\alpha} \log \left( \sum_{{c} \in \tilde{\mathcal{C}}} \Tr[\Gamma^c \omega] 2^{\frac{\alpha-1}{\alpha}f ({c})} \right. \nonumber \\
    & \qquad \qquad \left.+ \pkey \Tr[\mc{M}^\mathrm{K}(\omega)] 2^{\frac{1-\alpha}{\alpha} \left[-f ({\perp})+ H^\uparrow_\alpha (Z|E)_{\mc{M}(\omega)_{|{\perp}}}\right]}  \right). \label{eq:FWeightedDecomposition}
\end{align}
Where in the second line, we applied the conditioning on $C$ for the states, such that $Z=\bot$ whenever $C \neq \perp$, which effectively reduces the corresponding entropy to zero.

\subsection{Conic optimization}\label{subsec:ConicOptimization}

As a next step, we may apply a lower bound on the remaining Rényi entropy by using \cite{chung2025,navarro2025} 
\begin{align}
     H^\uparrow_\alpha(Z|{E})_{\mc{M}(\omega)_{|\bot}}  &\geq D_\gamma (\g(\omega_{ASR})\| \zg (\omega_{ASR})) \\
     &= \frac{-1}{1-\gamma } \log \left(\frac{\Psi_\gamma  \left(\g(\omega),\zg(\omega)\right)}{\Tr[\g(\omega)]}  \right), \label{eq:PsiBeta}
\end{align}
where $\gamma = 1/\alpha$, and $\g$, $\z$ are the key maps, which we explicitly define in Section \ref{Subsec:NumResults}. For now, we note that $\g$ is a trace-nonincreasing map that verifies $\Tr[\g (\omega)] = \Tr[\mc{M}^\mathrm{K}(\omega)]$. As a result, we introduce \eqref{eq:PsiBeta} in \eqref{eq:FWeightedDecomposition} to find 
\begin{align} 
    H^{\uparrow,f}_\alpha (Z|{C}E) \geq \frac{\alpha}{1-\alpha} \log \left(\sum_{{c} \in \tilde{\mathcal{C}}} \Tr[\Gamma^c \omega] 2^{\frac{\alpha-1}{\alpha}f ({c})} + \pkey 2^{\frac{\alpha-1}{\alpha} f ({\perp})} \Psi_\gamma \left(\g(\omega),\zg(\omega)\right) \right). \label{eq:BoundFWeighted}
\end{align}
Next, using the same procedures as in \cite[Eq. (solve for dual)]{kamin25MEATsecurity}, we formulate the fixed-length, honest implementation of the protocol to efficiently compute optimal values for the tradeoff function $f$. Namely, we ought to solve the minimization program for a given reference distribution $q$ defined on $C$ \cite{navarro2025}

\begin{equation}\label{eq:SecondMEATmain}
\begin{gathered}
    \min_{\omega, \lambda}   \frac{1}{1 - \gamma} D_\mathrm{KL}(\lambda ||{\mc{M}}(\omega)_{C}) +  q(\perp) D_{\gamma} (\g(\omega)\|\zg(\omega)) \\
    \mathrm{s.t.} \quad \omega_{ASR} \succeq 0, \, \Tr_S[\omega_{ASR}] = \sigma_{A} \otimes \omega_R, \\ 
    \;  \lambda - q = 0,
\end{gathered}
\end{equation}
and the dual values for the last line of constraints provide our choice for $f$. Given the aforementioned considerations, we can now explicitly reformulate the minimization defined by $f$ and $\kappa$ into conic programs. To this end, we employ the FastRényiQKD cone from \cite{navarro2025}

\begin{equation}\label{eq:FastCone}
    \mathcal{K}^\gamma = \left\{(u,\rho) \in \mathbb{R} \times \mathbb H^d_\succ; u \ge - \hat{\Psi}_\gamma(\rho) \right\},
\end{equation}
where we crucially note that  $\hat{\Psi}_\gamma (\cdot)$ originates from $\Psi_\gamma \left(\g(\cdot),\zg(\cdot)\right)$ after incorporating a facial reduction \cite{drusvyatskiy2017,hu2021robust} in the definition of the key maps $\g$ and $\z$ to ensure strict feasibility and a simplified numerical framework. Applying this construction in \eqref{eq:BoundFWeighted},  we find for \eqref{eq:KappaProp} the following bound\footnote{Note that the positive semidefiniteness is now incorporated in the definition of the FastRényiQKD cone.}

\begin{equation}\label{eq:Conic_kappaFast}
\begin{gathered}
    \kappa \geq  \min_{u, \omega} \frac{\alpha}{1-\alpha}  \log  \left(  \sum_{{c} \in \tilde{\mathcal{C}}} \Tr[\Gamma^c \omega] 2^{\frac{\alpha-1}{\alpha}f ({c})} - p^\mathrm{K} 2^{\frac{\alpha-1}{\alpha} f ({\perp})} u \right) \\ %
        \mathrm{s.t.} \quad \Tr_S[\omega_{ASR}] = \sigma_{A} \otimes \omega_R, \\ 
    (u,\omega) \in \mathcal{K}^{\gamma}.
\end{gathered}
\end{equation}
Since the negative logarithm is a monotonically decreasing function, we only need to perform a maximization for the argument of said logarithm, which is now given by an affine function and thus is amenable for standard conic programming.

Regarding the tradeoff function, we further use the well-known Kullback-Leibler divergence cone
\begin{equation}
    \mathcal{K}_\mathrm{KL} = \{(h_{\mathrm{KL}},q,p) \in \mathbb{R} \times \mathbb{R}_>^d \times \mathbb{R}_>^d ; h_{\mathrm{KL}} \geq D_\mathrm{KL}(q\|p)\}, 
\end{equation}
combined with a fixed choice $q(\bot) = p^\mathrm{K} \Tr[\mc{M}^{\mathcal{K}}(\omega)]$ for the reference, and the logarithmic cone 
\begin{equation}
    \mathcal{K}_{\log} = \left\{ (h, v, u) \in \mathbb{R} \times \mathbb{R}_> \times \mathbb{R}_> : h \leq v \log\left(u/v\right) \right\},
\end{equation}
which provides as a result a standard conic program from \eqref{eq:SecondMEATmain}

\begin{equation}\label{eq:Conic_2}
\begin{gathered}
    \min_{h_\mathrm{KL},h_\mathrm{QKD}, u, \lambda, \omega}  \frac{\alpha}{\alpha - 1} [h_\mathrm{KL} - p^\mathrm{K} h_\mathrm{QKD}] \\ %
    \mathrm{s.t.} \quad  \Tr_S[\omega_{ASR}] = \sigma_{A} \otimes \omega_R, \\ 
    (u,\omega) \in \mathcal{K}^{\gamma} \\
    \left(h_\mathrm{KL}, \lambda, \mathcal{{M}}(\omega)_C \right) \in \mathcal{K}_\mathrm{KL} \\
    \left(h_\mathrm{QKD},\Tr[\g(\omega)] ,- u \right) \in \mathcal{K}_{\log} \\
    q - \lambda = 0
\end{gathered}
\end{equation}

\section{Numerical results} \label{Subsec:NumResults}

Provided the security statement of Proposition \ref{Prop:MainProp}, we can quantify the expected secret key length distilled by RPSK given an $\varepsilon_\mathrm{PA}-$secure and $\varepsilon_\mathrm{EC}-$correct implementation. To this end, we take the expected value for the full tradeoff function and the error correction cost,
\begin{align}
    \mathbb{E}\!\left[\hat{f}(c_1^n) - \lambda_{\mathrm{EC}}(c_1^n) \right] &= \mathbb{E}\!\left[\hat{f}(c_1^n) \right] - \mathbb{E}\!\left[ \lambda_{\mathrm{EC}}(c_1^n)  \right] \\
    &=  \sum_{c \in \mc{C}} f(c) q(c)  -\lambda_{\mathrm{EC}}\!\left(q\right),
\end{align}
Namely, the expected value for the quantities that depend on $c_1^n$ converge to the case where the reference $q$ (corresponding to an implementation where Eve does not perform any attack) is observed. Therefore, $q$ is simply provided by an honest characterization of the channel. Using these mean values at \eqref{eq:SKR_Proposition}, we find the average secret key length  \cite{kamin25MEATsecurity}
\begin{align}
  \mathbb{E}\!\left[\ell \right] \ge n\bigl( f\cdot q +\kappa \bigr) -\lambda_{\mathrm{EC}}\!\left(q\right) - \left\lceil\log\frac{1}{\varepsilon_{\mathrm{EC}}}\right\rceil-\frac{\alpha}{\alpha-1}\log\frac{1}{\varepsilon_{\mathrm{PA}}}+2.
\end{align}
Within our context, we will take the values
\begin{align}
    \varepsilon_\mathrm{EC} = 10^{-11} &&  \varepsilon_\mathrm{PA} &= 9 \times 10^{-11} 
\end{align}
for the security tolerances. On the other hand, the key map $\g$ required to define the FastRényiQKD cone and calculate $\kappa$ in Subsection \ref{subsec:ConicOptimization} is provided by the superoperator\footnote{We note that the maps $\g$ and $\z$ actually act on an intermediate register and not the final key register $Z$. With a slight abuse of notation, we omit this technicality and defer to Ref. \cite{navarro2025} for a detailed description of the connection between such registers.}

\begin{equation}
    G = \sum_{x=0,1}\ket{x}_{Z} \otimes \pro{x}_A \otimes \sqrt{\bar{N}^{\top}_{SR}},
\end{equation}
which emerges from reformulating the key measurements at \eqref{eq:SingleRoundMap} as a quantum channel \cite{Lin2019Asymptotic}, according to Alice's measurements and Bob observing at least one click at the detectors. On the other hand, $\z$ is a standard pinching map provided by the superoperators $\{\pro{x}_{Z}\otimes \mathds{1}_{SR}\}_{x=0,1}$, which project the measurements from Alice (in the key generation rounds) into the final key register. 

However, $\g$ is not full rank as it expands the original matrix into a larger space, which leads to null eigenvalues and an ill-defined numerical optimization. As mentioned in Subsection \ref{subsec:ConicOptimization}, a facial reduction \cite{drusvyatskiy2017, hu2021robust} must be performed to ensure the strict feasibility of the final conic program \cite{lorente2024,navarro2025}. This process results in the new key superoperator  
\begin{equation}
    \tilde{G} = \mathds{1}_A \otimes \left(\mathds{1}_{SR} - \left[1- \sqrt{2p_d - p_d^2}\right] \pro{00}_{SR} \right).
\end{equation}
The absence of dark counts ($p_d = 0$), also reduces the support of Bob's register, creating a new nullspace. As a result, an extra facial reduction on his subspace is required for such scenario. This leads to the final key superoperator
\begin{equation}
    \tilde{G} = \mathds{1}_A \otimes (\ket{0}_{K}\bra{01}_{SR} + \ket{1}_{K}\bra{10}_{SR} + \ket{2}_{K}\bra{11}_{SR}),
\end{equation}
where $\{\ket{0}_{K},\ket{1}_{K},\ket{2}_{K}\}$ constitutes an orthonormal basis for a reduced subspace $K$. We finish the analysis of the secret key rate components by quantifying the error correction cost. This is provided by the length of the error correction register $\lambda_\mathrm{EC}(q)$, which is bounded by the Slepian-Wolf theorem \cite{SlepianWolf} 
\begin{align}
    \lambda_\mathrm{EC}(q) 
    &\leq f_\mathrm{EC} n H(Z|BI)
\end{align}
Here, $f_\mathrm{EC}\geq 1$ denotes the error correction efficiency, and the entropy can be analytically calculated provided the measurement outocomes of Alice and Bob\footnote{In particular, with the no-clicks which de facto constitute a postselection mechanism.}, as well as Alice's inputs. In our implementation, we set the value $f_\mathrm{EC} = 1.1$.

\subsection{Protocol simulation}

Let us now derive numerical results for our analysis by considering a quantum channel characterized by an optical fiber and imperfections in the preparation and detection of the states. As a consequence, diverse parameters must be included in the analysis, such as the excess noise at the fiber, the asymmetric modulation at Alice's laser, and dark counts for Bob's detection. We recall that the repetition rate of this implementation is limited (see Remark \ref{rem:enforceNS}) and emphasize that this example is purely for illustrational purposes, while noting that future work shall address the performance of RPSK for free-space channels, where said limitation can be avoided.

In an honest implementation, optical fibers constitute a phase invariant, additive white gaussian noise channel which results in an evolution for the initial coherent states into displaced thermal states
\begin{align}
    \pro{\beta} \to D(\sqrt{\eta}\beta)\rho_t(\delta) D^\dagger(\sqrt{\eta}\beta)
\end{align}
where $\eta$ represents the transmittance of the quantum channel and $\rho_t$ a thermal state with $\delta = 1+\xi \eta$ the total variance,  provided that $\xi$ represents the excess noise at Alice's side. Using the shorthand notation

\begin{align}
    \mc{I}_{m,n} = \bra{m}D(\sqrt{\eta}\beta)\rho_t(\delta) D^\dagger(\sqrt{\eta}\beta)\ket{n},
\end{align}
The probabilities can be calculated via the formula \cite[Appendix B]{LinTrusted2020}

\begin{align}
    \bra{m}D(\sqrt{\eta}\beta)\rho_t(\delta) D^\dagger(\sqrt{\eta}\beta)\ket{n} = \begin{cases}  \sqrt{\frac{m!}{n!}} \frac{\delta^m}{(1+\delta)^{n+1}} (\sqrt{\eta}\beta^*)^{n-m} e^{\frac{-\eta|\beta|^2}{1+ \delta}} L_m^{(n-m)}\left(\frac{-\eta|\beta|^2}{\delta (1+\delta)}\right), \, n \geq m \\
     \sqrt{\frac{n!}{m!}} \frac{\delta^n}{(1+\delta)^{m+1}} (\sqrt{\eta}\beta)^{m-n} e^{\frac{-\eta|\beta|^2}{1+ \delta}} L_n^{(m-n)}\left(\frac{-\eta|\beta|^2}{\delta (1+\delta)}\right), \, n < m 
    \end{cases}
\end{align}
where $L^{(j)}_k(\cdot)$ is the generalized, Laguerre's polynomial of order $j$ and degree $k$. In particular, we have the vacuum component
\begin{align}
    \bra{0}D(\sqrt{\eta}\beta)\rho_t(\delta) D^\dagger(\sqrt{\eta}\beta)\ket{0} &= \frac{1}{1 + \delta} e^{\frac{-\eta|\beta|^2}{1+ \delta}}.
\end{align}
In the case of correct clicks (normalized to parameter estimation rounds) with a dark count rate $p_d \in [0,1]$, we use the decomposition \eqref{eq:DarkCounts} to explicitly calculate the probabilities

\begin{align}
    \frac{q(\mathrm{CC})}{(1-\pkey)} &= \frac{1}{2} \Tr[\rho^{0}_{SR}\bar{M}^{0}] + \frac{1}{2} \Tr[\rho^{1}_{SR}\bar{M}^{1}] \\
    &= \frac{1}{4} \Tr[\rho^{0}_{SR}] - \frac{(1-p_d)^2}{4} \Tr[\rho^{0}_{SR}M^\bot] + \frac{1-p_d}{4} \Tr[\rho^{0}_{SR}(M^0 - {M}^{1})] \nonumber \\
    &\quad + \frac{1}{4} \Tr[\rho^{1}_{SR}] - \frac{(1-p_d)^2}{4} \Tr[\rho^{1}_{SR}M^\bot] - \frac{1-p_d}{4} \Tr[\rho^{0}_{SR}(M^0 - {M}^{1})] \\
     &= \frac{1}{2} - \frac{(1-p_d)^2}{2} \mc{I}^2_{0,0} + \frac{1-p_d}{2}\sum_{N=1}^\infty \sum_{m,n=0}^{N} \frac{N!\, \mc{I}_{m,n} \mc{I}_{N-m,N-n}}{2^N (N-m)! m! (N-n)!n!}[1 - (-1)^{m+n}]. \label{eq:p_CC}
\end{align}
Here we used the representation of the states in an honest implementation

\begin{align}
    \rho^{0/1}_{SR} = D(\sqrt{\eta}\beta)\rho_t(\delta) D^\dagger(\sqrt{\eta}\beta) \otimes D(\pm\sqrt{\eta}\beta)\rho_t(\delta) D^\dagger(\pm\sqrt{\eta}\beta), 
\end{align}
together with

\begin{align}
    \bra{m}D(-\sqrt{\eta}\beta)\rho_t(\delta) D^\dagger(-\sqrt{\eta}\beta)\ket{n} = - \mc{I}_{m,n},
\end{align}
which holds for $\beta \in \mathbb{R}$ whenever $m+n$ is odd, as it is the case of \eqref{eq:p_CC} as only said combination provides a nonzero value in the sums. Although there is no closed form to calculate the sum in $N$, it is possible to efficiently calculate its value up to floating point precision, as it is the case with any other relevant quantity in the numerical analysis. Similarly, wrong clicks can be computed and result in

\begin{align}
    \frac{q(\mathrm{WC})}{(1-\pkey)} &= \frac{1}{2} - \frac{(1-p_d)^2}{2} \mathcal{I}^2_{0,0} - \frac{1-p_d}{2}\sum_{N=1}^\infty \sum_{m,n=0}^{N} \frac{N!\, \mc{I}_{m,n} \mc{I}_{N-m,N-n}}{2^N (N-m)! m! (N-n)!n!}[1 - (-1)^{m+n}], \label{eq:p_WC}
\end{align}
and no-clicks constitute the closure of the probability set
\begin{align}
q(\mathrm{NC)} = (1-p_d)^2\mc{I}^2_{0,0}.
\end{align}

Given all these elements, we already have a complete description of the protocol, and we can proceed with the numerical estimation of secret key rates. To this end, we adapt the non-symmetric conic programming techniques provided in \cite{navarro2025}. This framework employs the optimization packages Hypatia \cite{coey2022performance,coey2022solving} and JuMP \cite{Lubin2023} in the programming language Julia \cite{JuliaLang}. 
\\

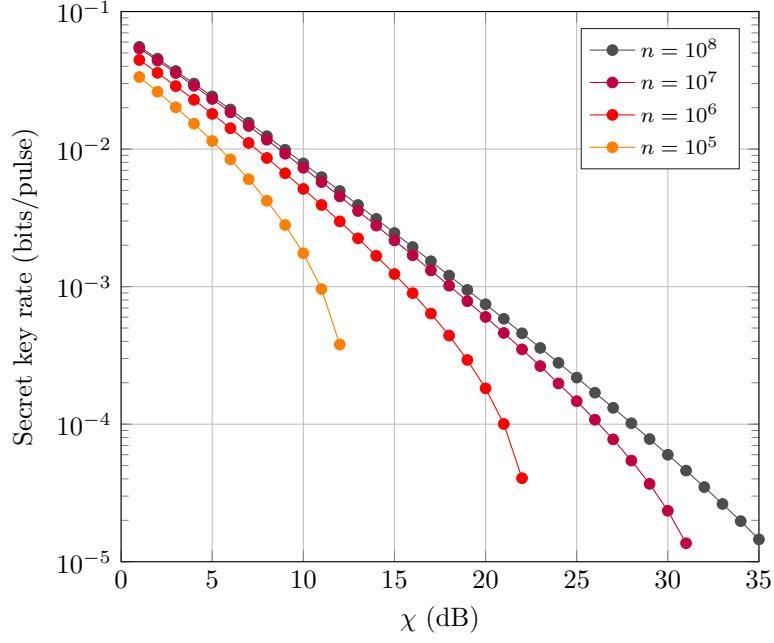
\begin{figure}[h!]
	\centering
 	\begin{tikzpicture}
		\begin{axis}[%
			scale only axis,
            ymode = log,
			xmin=0,
			xmax=35,
			ymin=1e-5,
			ymax=1e-1,
			grid=major,
			xlabel={$\chi$ (dB)},
            ylabel = {Secret key rate (bits/pulse)},
			axis background/.style={fill=white},
			legend style={at={(0.97,0.97)},legend cell align=left, align=left, draw=white!15!black, font=\footnotesize}
			]
            \addplot[black!70, mark=*] table[col sep=comma] {Plots/Fiber/FirstPlot/DPSK_Fiber_1e8.csv};
            \addlegendentry{$n = 10^{8}$};
            \addplot[purple, mark=*] table[col sep=comma] {Plots/Fiber/FirstPlot/DPSK_Fiber_1e7.csv};
        	\addlegendentry{$n = 10^{7}$}
            \addplot[red, mark=*] table[col sep=comma] {Plots/Fiber/FirstPlot/DPSK_Fiber_1e6.csv};
            \addlegendentry{$n = 10^{6}$}
            \addplot[orange, mark=*] table[col sep=comma] {Plots/Fiber/FirstPlot/DPSK_Fiber_1e5.csv};
        	\addlegendentry{$n = 10^5$ }
        \end{axis}
	\end{tikzpicture}
    \caption{Secret key generation rates for RPSK model under different block sizes $n$, $f_\mathrm{EC} = 1.10$, $\xi = 0.005$ and $p_d=0$. All other parameters were optimized for each point of the plot.} \label{fig:DPSK_Fiber}
\end{figure}

Figure \ref{fig:DPSK_Fiber} provides the secret key rate that can be distilled under no dark counts and different block sizes according to the channel loss in decibels $\chi$ (i.e., such that the transmittance verifies $\eta = 10^{-\chi/10}$). Although this plot is not fully comparable with the one provided in \cite[Fig. 8]{Sandfuchs2025} since said Ref. uses a mor conservative bound for the error correction cost, we observe a clear improvement in the block size as we are able to achieve losses of $\chi=$ 30 dB with only $n=10^7$, whereas \cite{Sandfuchs2025} needs much larger block sizes to distill the secret key.

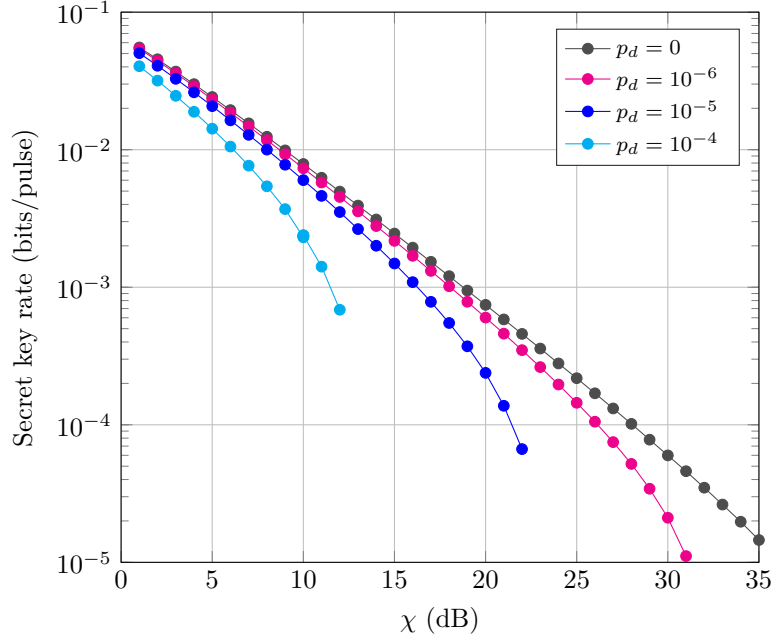
\begin{figure}[h!]
	\centering
 	\begin{tikzpicture}
		\begin{axis}[%
			scale only axis,
            ymode = log,
			xmin=0,
			xmax=35,
			ymin=1e-5,
			ymax=1e-1,
			grid=major,
			xlabel={$\chi$ (dB)},
            ylabel = {Secret key rate (bits/pulse)},
			axis background/.style={fill=white},
			legend style={at={(0.97,0.97)},legend cell align=left, align=left, draw=white!15!black, font=\footnotesize}
			]
            \addplot[black!70, mark=*] table[col sep=comma] {Plots/Fiber/FirstPlot/DPSK_Fiber_1e8.csv};
            \addlegendentry{$p_d = 0$};
            \addplot[magenta, mark=*] table[col sep=comma] {Plots/Fiber/DarkCount/DPSK_Fiber_pd1e6.csv};
        	\addlegendentry{$p_d = 10^{-6}$}
            \addplot[blue, mark=*] table[col sep=comma] {Plots/Fiber/DarkCount/DPSK_Fiber_pd1e5.csv};
            \addlegendentry{$p_d = 10^{-5}$}
            \addplot[cyan, mark=*] table[col sep=comma] {Plots/Fiber/DarkCount/DPSK_Fiber_pd1e4.csv};
        	\addlegendentry{$p_d = 10^{-4}$ }
        \end{axis}
	\end{tikzpicture}
    \caption{Secret key generation rates for RPSK under different dark count rates $p_d$, $f_\mathrm{EC} = 1.10$, $\xi = 0.005$ and $n=10^{8}$. All other parameters were optimized for each point of the plot.} \label{fig:DPSK_DarkCount}
\end{figure}

\newpage

\subsection{Analysis of imperfections}

Let us now focus on scenarios where practical imperfections are present in the  description and characterization of the protocol. As a first step, we evaluate the impact of dark counts on the key rate. Figure \ref{fig:DPSK_DarkCount} provides secret key rate curves for $n = 10^8$ rounds and different values of $p_d$. As expected, dark counts generally do not affect the secret key length until the transmittance losses set signal clicks to a low rate -- such that the random clicks (due to dark counts, which remain constant with respect to the transmittance) dominate Bob's measurement outcomes \cite{Boileau_2005}.

 As a result, Bob's observations become uncorrelated with respect to Alice, decreasing the key length to zero.

\begin{figure}[h!]
	\centering
 	\begin{tikzpicture}
		\begin{axis}[%
			scale only axis,
            xmode = log,
			xmin=5e-6,
			xmax=5e-1,
			ymin=2e-3,
			ymax=1e-2,
			grid=major,
			xlabel={$\alpha-1$},
            ylabel = {Secret key rate (bits/pulse)},
			axis background/.style={fill=white},
			legend style={at={(0.97,0.97)},legend cell align=left, align=left, draw=white!15!black, font=\footnotesize}
			]
            \addplot[black!70, mark=*] table[col sep=comma] {Plots/Fiber/DPSK_Fiber_alp1e9.csv};
            \addlegendentry{$n = 10^9$};
            \addplot[purple, mark=*] table[col sep=comma] {Plots/Fiber/DPSK_Fiber_alp1e8.csv};
        	\addlegendentry{$n = 10^{8}$}
            \addplot[red, mark=*] table[col sep=comma] {Plots/Fiber/DPSK_Fiber_alp1e7.csv};
            \addlegendentry{$n = 10^{7}$}
            \addplot[orange, mark=*] table[col sep=comma] {Plots/Fiber/DPSK_Fiber_alp1e6.csv};
        	\addlegendentry{$n = 10^{6}$ }
        \end{axis}
	\end{tikzpicture}
    \caption{Secret key generation rates for RPSK under values for the Rényi parameter $\alpha$, $p_d = 0$, $f_\mathrm{EC} = 1.10$, $\chi = 10$ dB, $\xi = 0.005$, $\pkey = 0.96$ and diverse block sizes $n$. The amplitude of the coherent states was set to $\beta = 0.45$.} \label{fig:DPSK_Alphas}
\end{figure}
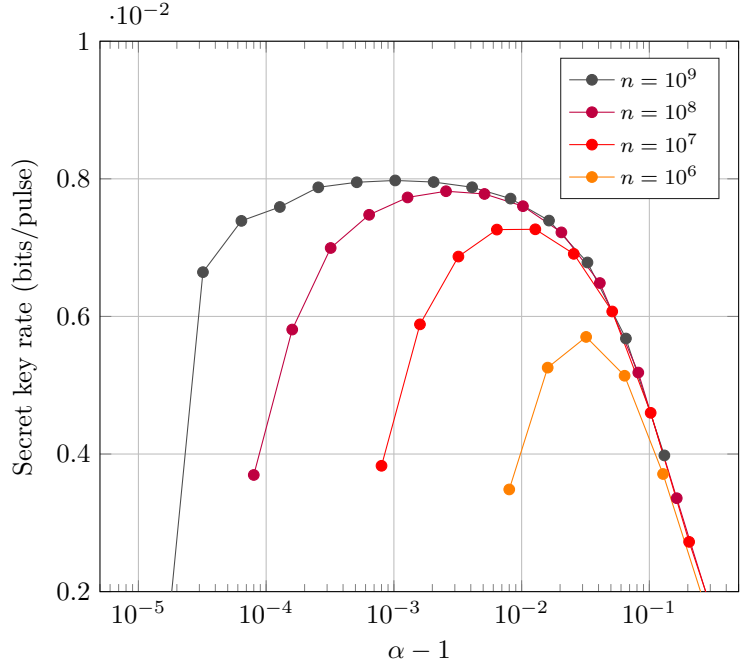

On the other hand, achieving an optimal, variable-length secret key relies on an appropriate choice for the Rényi parameter $\alpha$. In this respect, Figure \ref{fig:DPSK_Alphas} provides the secret key that Alice and Bob observe for different block sizes according to $\alpha$, with fixed values $\pkey = 0.96$, $\beta = 0.45$, $\chi = 10$ dB and no dark counts. As we can see, diminishing block sizes require a sharper optimization for the Rényi parameter. Within the context of conic programming, this task can be efficiently done by adding unidimensional line search routines (such as Nelder-Mead \cite{nocedal2006numerical}) that perform said search. We do not observe an overhead cost by adding said routine, as solving conic programs for reduced dimensions typically takes a time in the order of miliseconds \cite[Table 1]{navarro2025}. 

\begin{figure}[h!]
	\centering
 	\begin{tikzpicture}
		\begin{axis}[%
			scale only axis,
            ymode = log,
			xmin=0,
			xmax=25,
			ymin=1e-5,
			ymax=1e-1,
			grid=major,
			xlabel={$\chi$ (dB)},
            ylabel = {Secret key rate (bits/pulse)},
			axis background/.style={fill=white},
			legend style={at={(0.97,0.97)},legend cell align=left, align=left, draw=white!15!black, font=\footnotesize}
			]
            \addplot[black!70, mark=*] table[col sep=comma] {Plots/Fiber/gammas/DPSK_Fiber_gamma1.csv};
            \addlegendentry{$\mu = 1$};
            \addplot[magenta, mark=*] table[col sep=comma] {Plots/Fiber/gammas/DPSK_Fiber_gamma11.csv};
        	\addlegendentry{$\mu = 1.1$}
            \addplot[blue, mark=*] table[col sep=comma] {Plots/Fiber/gammas/DPSK_Fiber_gamma115.csv};
            \addlegendentry{$\mu = 1.15$}
            \addplot[cyan, mark=*] table[col sep=comma] {Plots/Fiber/gammas/DPSK_Fiber_gamma12.csv};
        	\addlegendentry{$\mu = 1.2$ }
        \end{axis}
	\end{tikzpicture}
    \caption{Secret key generation rates for RPSK under different asymmetry factors $\mu$, coherent state amplitude $\beta = 0.45$, $p_d = 10^{-5}$, $f_\mathrm{EC} = 1.10$, $\xi = 0.005$ and $n=10^{7}$. The Rényi parameter $\alpha$ and $\pkey$ were optimized for each point of the plot.} \label{fig:DPSK_gamma}
\end{figure}
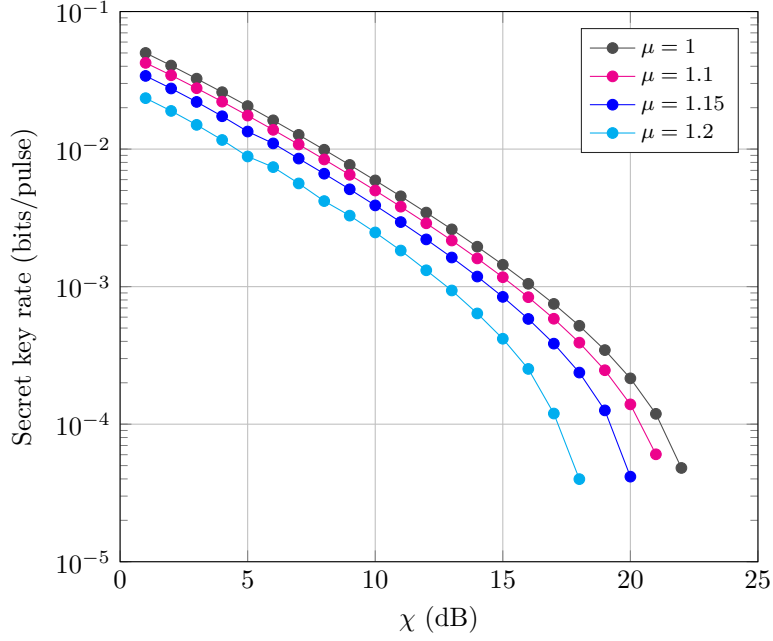

We conclude the analysis by considering the robustness of the protocol against an imperfect modulation by Alice, such that the amplitude of her coherent states is actually $(-\beta,\mu \beta) $ with $\beta$ the ideal amplitude and $\mu$ an asymmetry factor. Due to the structure of the protocol, $(-\beta,\mu \beta) $ yields the same results as $( -\mu\beta, \beta)$. As Figure \ref{fig:DPSK_gamma} shows, where $\beta = 0.45$, $p_d = 10^{-5}$, and $n=10^{7}$ are taken, the protocol performance is robust against these deviations, which only result in a clear damage for the key rate when $\mu > 1.2$, a factor well within the range of experimental implementations. The only neat change in terms of protocol performance lies in the optimization with respect to the Rényi parameter, as the line search increasingly becomes unstable when the asymmetry factor increases. In spite of these imperfections, we still observe that secret keys can be achieved beyond 15 dB of transmittance loss.

\section{Discussion} \label{Sec:Discussion}


With this work, we have expanded the security proof for RPSK from Ref. \cite{Sandfuchs2025} by formulating a variable-length, composable general security proof via the MEAT that also incorporates an analysis of imperfections.  Further improvements include lifting previous repetition rate limitations inherited from the structure of the GEAT \cite{metger2022security}, and a study of the optimal protocol performance by adjusting diverse parameters (such as the key round probability, or the amplitude of the coherent states), particularly under the presence of experimental imperfections.

These steps lead to a consistent reduction of orders of magnitude in the block size compared to previously reported findings \cite{Sandfuchs2025, Mitzutani2024}. In particular, we achieve secret key rates for  $n = 10^5$ rounds beyond a loss of 12 dB, which comes into huge constrast with the values $n \approx 10^{12}$ from Ref. \cite{Mitzutani2024}. Our analysis also enhances the overall security standards reported for RPSK, with an extension of the total tolerance parameter for the error correction validation to $\varepsilon_\mathrm{EC} = 10^{-11}$ without compromising the final protocol yield. This value is in line with the usual order of magnitude $\varepsilon \approx 10^{-10}$, which is again a clear improvement over the value $\varepsilon_\mathrm{EC} = 10^{-2}$ from Ref. \cite{Sandfuchs2025}.

Furthermore, we have generalized the original squashing map by adding the dark count model as outlined in Ref. \cite{Nahar_2026}, and studied the effect of symmetric dark counts on the secret key rate. As we noted in Section \ref{Sec:ProtocolDescription}, Bob can enforce said symmetry by adding artificial noise at his setup. However, it would be of clear interest to add different dark counts for each detector, as well as other trusted imperfections. These may include an unbalanced beam splitter, depolarizing noise at Alice's states, imperfect randomness sources, or adding considerations of asymmetric authentication \cite{ferradini2025definingsecurityquantumkey}.

With respect to the numerical method, we note that the efficient conic optimization from Ref. \cite{navarro2025} allows the computation of secret key rates in real time, including a line search optimization for $\alpha$. This constitutes a clear advance as it allows the characterization of the key length during a QKD session without generating a bottleneck, or requiring an excessively precise characterization of the quantum channel -- which often causes an abortion of the protocol \cite{Tupkary2024,Kanitschar_2025}.

These advances, however, come with a caveat as indicated in Remark \ref{rem:enforceNS} -- the no-signaling constraints required to efficiently perform the protocol represent again a limitation on the repetition rate of the protocol according to the total distance between Alice and Bob. However,  and on the contrary of the GEAT, this is not a fundamental requirement of the security proof, and depends only on the speed of light at the quantum channel. Therefore, it represents a complication for fiber-based systems, but not for free-space propagation, where the speed of light is nearly equal to $c$. 

On a different note, it is still unclear whether the MEAT and similar techniques be can used to extend the findings herewith provided to the case of DPSK due to the sequential structure of the latter. Future work shall address whether EAT variants that make explicit use of sequential structures, such as the generalized Rényi EAT \cite{arqand2024generalized}, can be exploited to improve the finite-size corrections, and possibly lift repetition rate limitations via memory registers.

To summarize, we have proved that RPSK is within accessible range for experimental demonstrations thanks to the reduction of the minimum block size, allowing an implementation of QKD with commercial devices. Further research shall address the feasibility and experimental approach of the protocol herewith analyzed, particularly for satellite and free-space links where the protocol constitutes a clear pathway towards practical QKD. 

\section*{Acknowledgments}
I thank Devashish Tupkary for insightful indications about the limitations of the MEAT framework for the security of DPSK. This project has received funding from the European Union’s Digital Europe Programme under the project QUARTER (101091588), and from the European Innovation Council's Horizon Europe EIC Accelerator Programme under the project MIQRO (101161539), and the European Union (QSNP, 101114043).

\bibliographystyle{alpha}
\bibliography{CV}
\end{document}